\documentclass[final,5p,times,twocolumn,authoryear]{elsarticle}

\usepackage{amssymb}

\usepackage{float}
\usepackage{threeparttable}

\journal{Computerized Medical Imaging and Graphics}

\begin{document}

\begin{frontmatter}



\title{DXM-TransFuse U-net: Dual Cross-Modal Transformer Fusion U-net for Automated Nerve Identification}


\author[inst1]{Baijun Xie}

\author[inst1]{Gary Milam}

\affiliation[inst1]{organization={Department of Biomedical Engineering, School of Engineering and Applied Science, The George Washington University},
            addressline={800 22nd Street NW, 5000 Science \& Engineering Hall}, 
            city={Washington DC},
            postcode={20052}, 
            state={DC},
            country={U.S.A.}}

\author[inst2]{Bo Ning}
\author[inst2]{Jaepyeong Cha}

\affiliation[inst2]{organization={Children’s National Hospital},
            addressline={111 Michigan Avenue NW}, 
            city={Washington DC},
            postcode={20010}, 
            state={DC},
            country={U.S.A.}}

\author[inst1]{Chung Hyuk Park$^*$}

\begin{abstract}
Accurate nerve identification is critical during surgical procedures for preventing any damages to nerve tissues. Nerve injuries can lead to long-term detrimental effects for patients as well as financial overburdens. In this study, we develop a deep-learning network framework using the U-Net architecture with a Transformer block based fusion module at the bottleneck to identify nerve tissues from a multi-modal optical imaging system. By leveraging and extracting the feature maps of each modality independently and using each modalities information for cross-modal interactions, we aim to provide a solution that would further increase the effectiveness of the imaging systems for enabling the noninvasive intraoperative nerve identification. 
\end{abstract}




\end{frontmatter}


\section{INTRODUCTION}
Neurogenic injury as a result of surgical mishap can present significant morbidity and even mortality depending on the location of injury~\citep{antoniadis2014iatrogenic,kretschmer2009iatrogenic,rasulic2017iatrogenic}. Injuries to motor neurons, in particular, can have critical impacts on the quality of life. A facial nerve injury during a head and neck surgery results in facial paralysis that includes asymmetry of facial expressions, difficulties in eating or drinking, loss of blinking control, and roping of the mouth on the affected side~\citep{gordin2015facial}. Damages to the recurrent laryngeal nerve (RLN) during thyroidectomy, open neck, or cardiac surgery could induce paresis or palsy of the vocal cord with voice or swallowing dysfunction. Likewise, bilateral RLN injury could result in airway obstruction, resulting in the requirement for a tracheostomy~\citep{snyder2008elucidating}. A pelvic nerve injury after a prostatectomy or rectal cancer surgery is associated with urinary and sexual dysfunctions~\citep{bohrer2009pelvic}. Finally, spinal cord surgery, as in the release of a tethered cord, can have dire consequences on bladder, bowel, or lower extremity motor function if the inadvertent neural injury occurs~\citep{duz2008tethered}. The variety of postoperative nerve injuries described persisting even in the most experienced hands. Thus, there is an unmet clinical need for a noninvasive intraoperative nerve identification to increase situational awareness during operation.

As a standard of care, visual inspections have been performed by operating surgeons that rely on individual surgeon’s experience and training. More recently, intraoperative neuromonitoring devices have been introduced and adopted for nerve localization~\citep{cha2018real}. However, this technique requires intermittent electrical stimulations via an electrode probe to confirm the neuromuscular activity, which can interrupt surgical workflows. As an alternative, our group recently proposed and successfully demonstrated an optical imaging nerve identification using the Mueller polarimetric imaging~\citep{ning2021improved}, which calculates intrinsic birefringence patterns from fibrous nerve structures. In our previous study~\citep{ning2021improved}, the system showed a promise but had a limitation of nerve-specific segmentation due to the presence of similar birefringence signals from other surrounding fibrous anatomies of tendons and collagenous muscle tissues. Therefore, we concluded that more advanced image processing with larger imaging data would further improve the specificity of the optical nerve identification.   

In recent years, deep convolution neural networks (CNNs) have shown state-of-the-art results in many computer vision and medical imaging tasks. CNN has achieved remarkable results in the task of image classification~\citep{krizhevsky2012imagenet}. In addition to image classification, CNN has been designed to solve semantic segmentation tasks as well. \citet{ciresan2012deep} utilized a sliding window technique to train the segmentation CNN. A fully-connected CNN is also proposed to achieve end-to-end semantic segmentations~\citep{long2015fully}. 

Despite the breakthrough of CNN in the computer vision and medical imaging domains, one drawback is that CNN structures usually require a massive amount of data to train on. However, in the field of medical imaging, the acquisition process of images involves many layers of medical protocols and privacy/security issues, on top of which experts' annotations are also heavily required. These make the usage of CNN more expensive and time-consuming, even if the data becomes accessible. In a recent study, the U-Net~\citep{ronneberger2015u} model alleviates this problem by using an encoder-decoder architecture. The encoder-decoder structure and the so-called ``skip connection'' enable the U-Net to get more prominent performance with a small amount of training data. Recent studies have also made improvements based on the U-Net and shown better performances in the semantic segmentation tasks. UNet++ consist of nested and dense skip connections to get more effective results~\citep{zhou2018unet++}. \citet{oktay2018attention} proposed the network with an attention gate (AG) module at the decoder part of the U-Net, which targets to learn the salient features for a specific task. The results showed better prediction performance on different datasets while preserving computational efficiency.

Current research efforts have also shown the effectiveness of employing U-Net to tackle the semantic segmentation task in the field of medical imaging. Ultrasound nerve segmentation images dataset from Kaggle competition has been widely used and trained by U-Net in many studies~\citep{baby2017automatic, kakade2018identification, rubasinghe2019ultrasound}. The possibility of the adaptation of the generative adversarial network (GAN) has also been discussed~\citep{rubasinghe2019ultrasound}. Moreover, U-Net has been used to identify more complicated nerve structures from ultrasound images that contain the musculocutaneous, median, ulnar, and radial nerves, as well as the blood vessels~\citep{smistad2018highlighting}. Other studies have also deployed the U-Net model to identify other nerve structures, such as the proposed corneal nerve segmentation network~\citep{wei2020deep}, for sub-basal corneal nerve segmentation and evaluation. \citet{yamato2020nerve} have also shown that U-Net can be applied to the nerve segmentation of coherent anti-Stokes Raman scattering endoscopic images.

\section{RELATED STUDIES}
More recent studies are exploring the possibility of using U-Net based structure to combine different image modalities for training~\citep{dolz2018dense,dolz2018ivd,Kumar_2020}. A multi-path architecture is employed that can extract and combine the unique features from different modalities~\citep{dolz2018dense,dolz2018ivd}. \citet{Kumar_2020} have presented a new CNN that aims to fuse complementary information in multi-modality images for medical image segmentation. The strength of that study is introducing the co-learning component that allows the model to learn the fusion of different modalities and the importance of a specific modality by placing weight accordingly to the feature maps of each modality.

Due to the fact that salient parts in medical imaging can vary greatly, \citet{dolz2018dense, dolz2018ivd} proposed an extended inception module that would be able to learn from different receptive fields. Using convolutions of multiple kernel sizes at the same level allows capturing both local and general information; furthermore, two additionally dilated convolution blocks with different dilated rates were leveraged to help increase the captured global context. Although the architectures proposed by~\citet{dolz2018dense, dolz2018ivd} leverages extended inception modules and dense connectivity between multiple paths of the encoder to show improvements over other fusion techniques, the increase in model size and the potential delay in the inference time that comes along with those models are still existing limitations.

Therefore, our idea is that leveraging attention mechanisms to focus on the region of interest and learning the importance of the higher-level features of each modality would increase the model's prediction capabilities over other single U-Net variants while maintaining the size and inference time at an acceptable range. A recent study shows that the attention mechanism purely based on multi-head attention~\citep{vaswani2017attention} has significant advantages, which has also been applied to the task of medical image segmentation~\citep{chen2021transunet, petit2021u}. \citet{chen2021transunet} proposed a model which took advantage of both the U-Net and the Transformer and showed that Transformer could act as a strong encoder in U-Net architecture. A study proposed by~\citet{petit2021u} also added the multi-head attention module at each skip connection and demonstrated better performance than the previous study of attention gates mechanism~\citep{oktay2018attention}. However, existing studies have only focused on applying Transformer for U-Net on the unimodal dataset. To the best of the authors' knowledge, few studies have used Transformer for multi-modal fusion in the task of medical image segmentation. The usage of Transformer for cross-modal interaction has been proposed and applied mainly in the study of natural language processing~\citep{tsai2019multimodal}.

Thus, our study aims to improve the performance of the intraoperative nerve identification system by using multi-modal medical images with deep neural networks. In this study, we present an improved Dual Cross-Modal Transformer Fusion U-net (DXM-TransFuse U-net), which uses a multi-path architecture that learns to fuse different image modalities via cross-modal interactions, where each modality of the image is treated as an independent signal. Furthermore, a Transformer module that was inspired by \citet{tsai2019multimodal} is implemented at the bottleneck for the fusion of different modalities. The significance of our contribution lies in designing the enhanced U-Net architecture and analyzing and comparing the performance of the different combinations of image modalities with real nerve tissue data collected during intraoperative processes. In addition, we conduct experiments to demonstrate that our proposed network is more efficient compared with other baseline models.



\section{MATERIALS AND METHODS}

\subsection{Animal Procedures \& Dataset Preparation}
Sample nerve images were acquired at the Children’s National Research Animal Facility under the approval of the Institutional Animal Care and Use Committee (IACUC \#30591). After euthanasia of the living animals (N=4), the cervical incisions on the central neck of each pig were performed using standard surgical instruments: sharp dissection by scalpel, blunt dissection by scissors/forceps, and coagulation by electrocoagulation. The ventral portion of the superficial neck muscle was exposed. Vagus nerves, recurrent laryngeal nerves, and superior laryngeal nerves were dissected and targeted for imaging using an in-house dual-RGB/polarimetric imaging system~\citep{ning2021improved} similar to the previous work~\citep{cha2018real}.

\subsection{Birefringence Images}

The birefringence images are derived from a polarimetric camera using the Mueller matrix decomposition, described in our earlier work~\citep{cha2018real}.

\begin{figure}[thpb]
    \centering
	\includegraphics[scale=0.28]{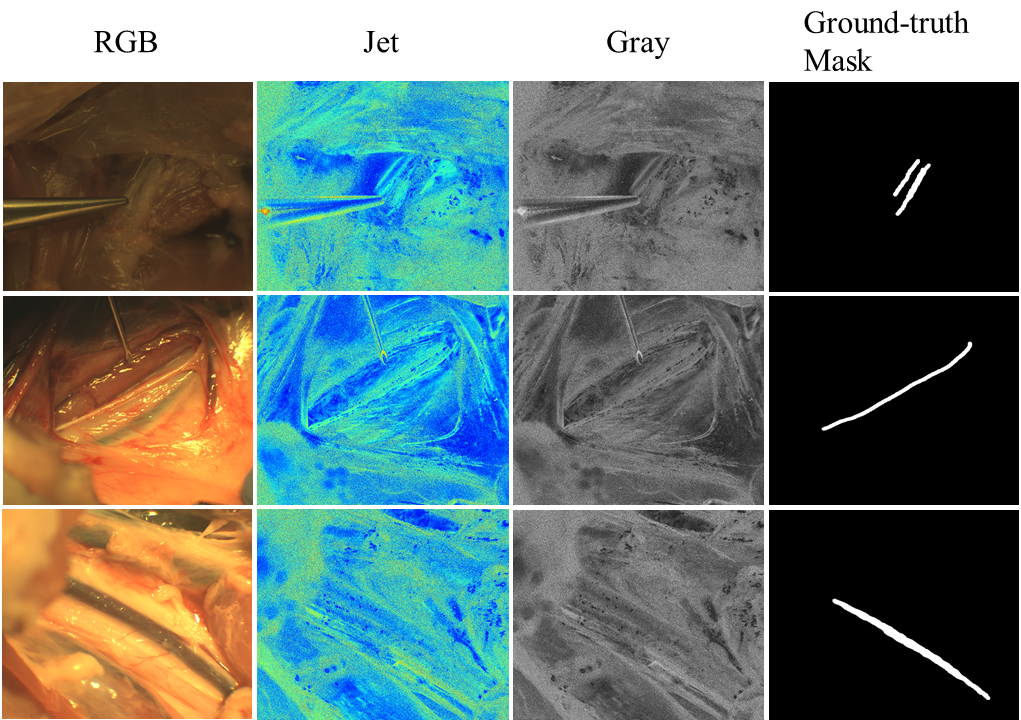}
	\caption{Samples of the generated images and masks.}
	\label{fig:samples}
\end{figure}

The multi-modal optical imaging system provides an RGB and birefringence outputs (Jet and Grayscale), overall consisting of three types of images: RGB, Jet, and Grayscale. As shown in Figure~\ref{fig:samples}, the first three columns display the three image types with different modalities, with the last column displaying the ground-truth mask annotated by the expert. As the two birefringence image types are of the same source, we would only be using one type for our models, with the Jet representation being chosen as it consists of 3 channels and carries more information that would be significant during training.

\subsection{Expert Mask Annotations}
Ground-truth masks were annotated by two experienced surgeons (Drs. Sandler and Kim) from Children’s National Hospital after the confirmation of Neuromonitoring device (Nerveana Nerve Locator and Monitor, NeuroVision Medical Products, Ventura, CA). The last column of Figure~\ref{fig:samples} shows the samples of expert annotated masks in this study. A user-oriented masking software was used. The marker-based watershed algorithm provided by the OpenCV library~\citep{opencv_library} was implemented on the software to provide the expert users with the option to utilize the watershed to assist in the mask creation.

\subsection{Datasets} \label{datasets}

The used dataset is curated to ensure images from all modalities are in line with each other as this would be essential for the multi-modal network approach. The entire dataset consists of 188 images of each modality: birefringence (Jet) and RGB images.

The dataset is then divided into 5 groups for establishing the 5-fold cross-validation procedure, which is desired in the absence of a large dataset. For preprocessing, images are resized to $256\times256$, with standardization and normalization techniques applied. Normalization is performed using the channel means and standard deviations of their respective image types.

 \begin{figure}[htbp]
    \centering
	\includegraphics[scale=0.28]{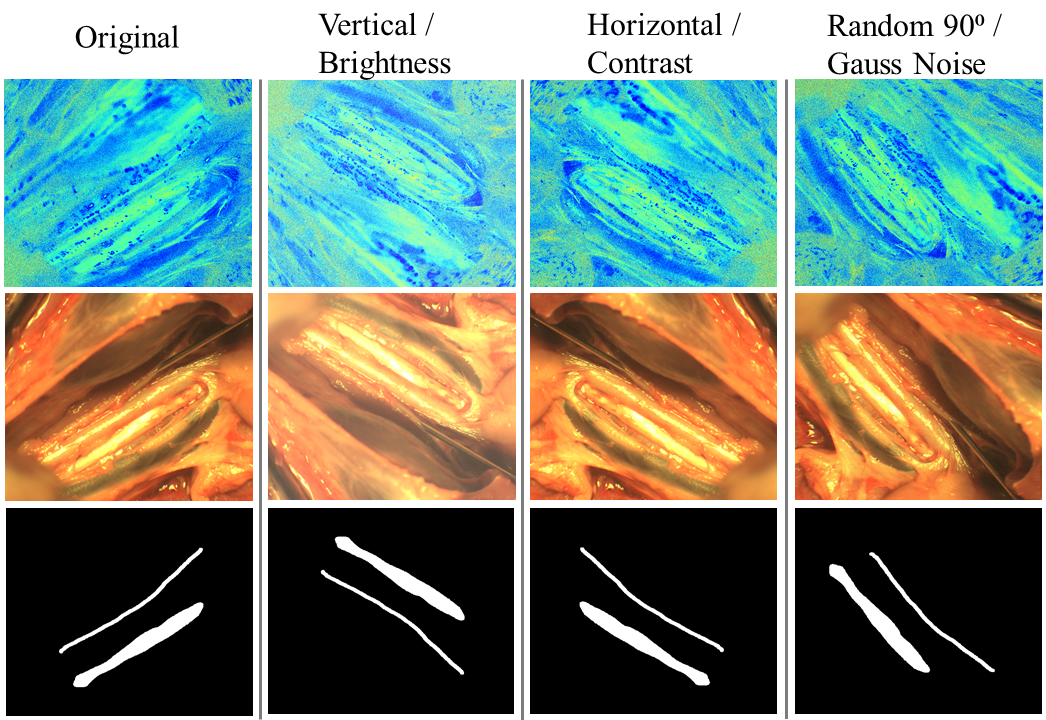}
	\caption{Samples of augmentation images and the corresponding masks.}
	\label{fig:augmentation}
\end{figure}

Positional and color augmentation techniques are performed on each image of the training set and are added to the set. Training images are vertically flipped with the brightness randomly modified, horizontally flipped with contrast randomly modified, rotated at a randomly selected degree of multiples of 90 degree, and applied with Gaussian noise; only positional augmentations are performed on the respective masks. Examples of augmented images and masks are shown in Figure~\ref{fig:augmentation}.

\subsection{Deep Neural Network Architecture}

To demonstrate the effectiveness of the proposed multi-modal deep neural network, we first compare the proposed architecture to the single modality network. For the single modality, U-Net~\citep{ronneberger2015u} and Attention U-Net~\citep{oktay2018attention} are considered as the baseline of the study. Original U-Net contains an encoder and a decoder, which is developed for biomedical image segmentation. The encoder part is called the contraction path, which consecutive stacking with convolutions followed by max-pooling layers for down-sampling. The decoder part is the expansion path which is up-sampling by up-convolution. To address the problem of losing information during the up-sampling, the U-Net uses a skip connection, which integrates the information from the contraction path. 

\begin{figure}[thpb]
    \centering
	\includegraphics[scale=0.27]{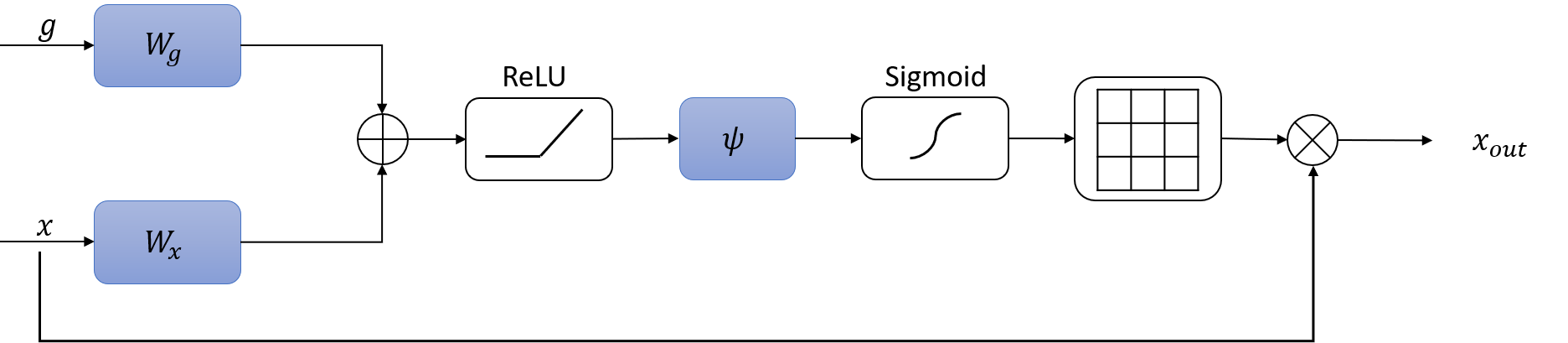}
	\caption{Schematic of the additive AG.}
	\label{fig:AG}
\end{figure}

Attention U-Net differs from U-Net by the addition of soft attention gates at the skip connections~\citep{oktay2018attention}. The attention mechanism enables the network to focus on the region of interest and reduces redundant features. As can be seen in Figure~\ref{fig:AG}, $x_l$ is the output feature map from the previous layer. Gating signal $g$ is collected from a coarser scale that provides contextual information for the focus spatial regions. The attention coefficient, $\alpha$, is used to identify salient features and suppress redundant features to reserve the activations relevant to the specific task. The attention coefficient is formulated as follows:
\begin{equation}
    \alpha  = \sigma_2 (\psi^{\intercal}(\sigma_1(W_x^{\intercal}x_l + W_g^{\intercal}g + b_g)) + b_{\psi})
\end{equation}
 
\noindent where $\sigma_1$ is the ReLU function, $\sigma_2$ is the Sigmoid function, $W_x$, $W_g$, and $\psi$ are linear transformations, and $b_g$ and $b_{\psi}$ are the bias terms.

The output of AGs $x_{out}$ is the element-wise multiplication of input feature map $x_l$ and the attention coefficient $ \alpha$:

\begin{equation}
    x_{out} = x_l \cdot \alpha
\end{equation}

The Cross-attention module introduced in ~\citep{petit2021u} is investigated as well as another attention mechanism for the skip connections. Similar to the Attention U-net, it is used to put focus on the regions-of-interests (ROI) while suppressing irrelevant information. The module uses a multi-head attention module that uses as inputs the higher level feature maps from the previous layer and the skip connection information. A sigmoid activation is then applied on the computed values, which is then followed by an element-wise multiplication of it with the skip connection.

Additionally, the transposed convolutions on the decoder path had been substituted with a bicubic upsampling followed by a convolution operation. Although the bilinear or bicubic upsamplings are more expensive than transposed convolutions, they have shown to suffer less from artifacts~\citep{47862}. The bicubic method is chosen over the bilinear because sixteen nearest neighbors are used on the former instead of only four nearest neighbors on the latter. In addition, leveraging more information during the upsampling will produce better segmentation results.

\subsection{Multi-Modal Deep Neural Network Architecture}

Two fusion approaches are investigated for the multi-modal baseline comparison. One is a late-concatenation fusion approach at the bottleneck of the U-Net and the other is adapting the co-learn module proposed by \citet{Kumar_2020} at the bottleneck. The late-fusion approach was chosen over early-fusion as it is seen to outperform early fusion on previous studies, such as this study by \citet{dolz2018ivd}. The co-learn module is implemented at the bottleneck as oppose to each layer to allow a fair comparison across all three fusion approaches and maintain the model at a reasonable size.

\begin{figure}[thpb]
    \centering
	\includegraphics[scale=0.6]{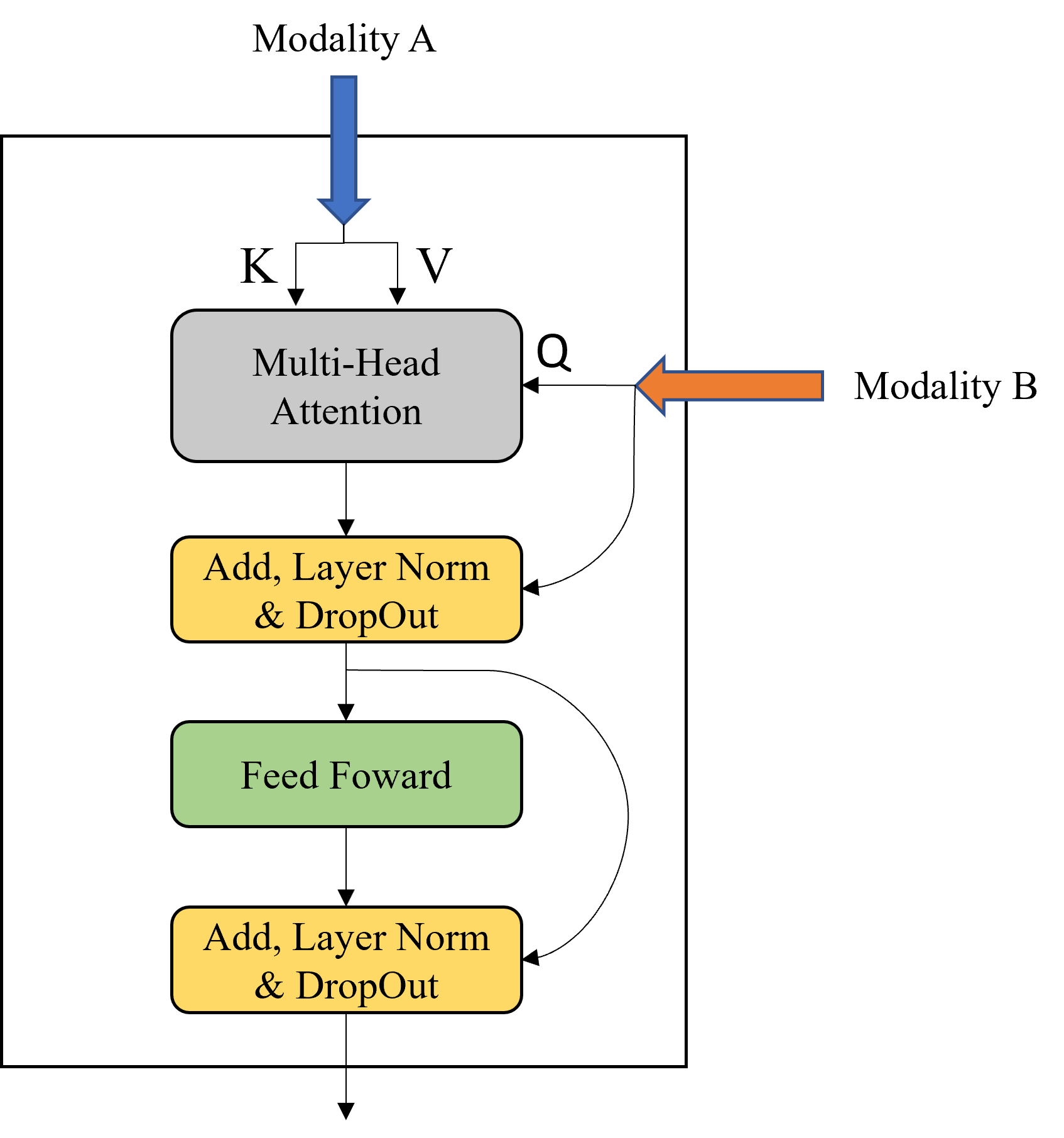}
	\caption{The cross-modal Transformer block.}
	\label{fig:crossmodal}
\end{figure}
\begin{figure*}[thpb]
    \centering
	\includegraphics[scale=0.54]{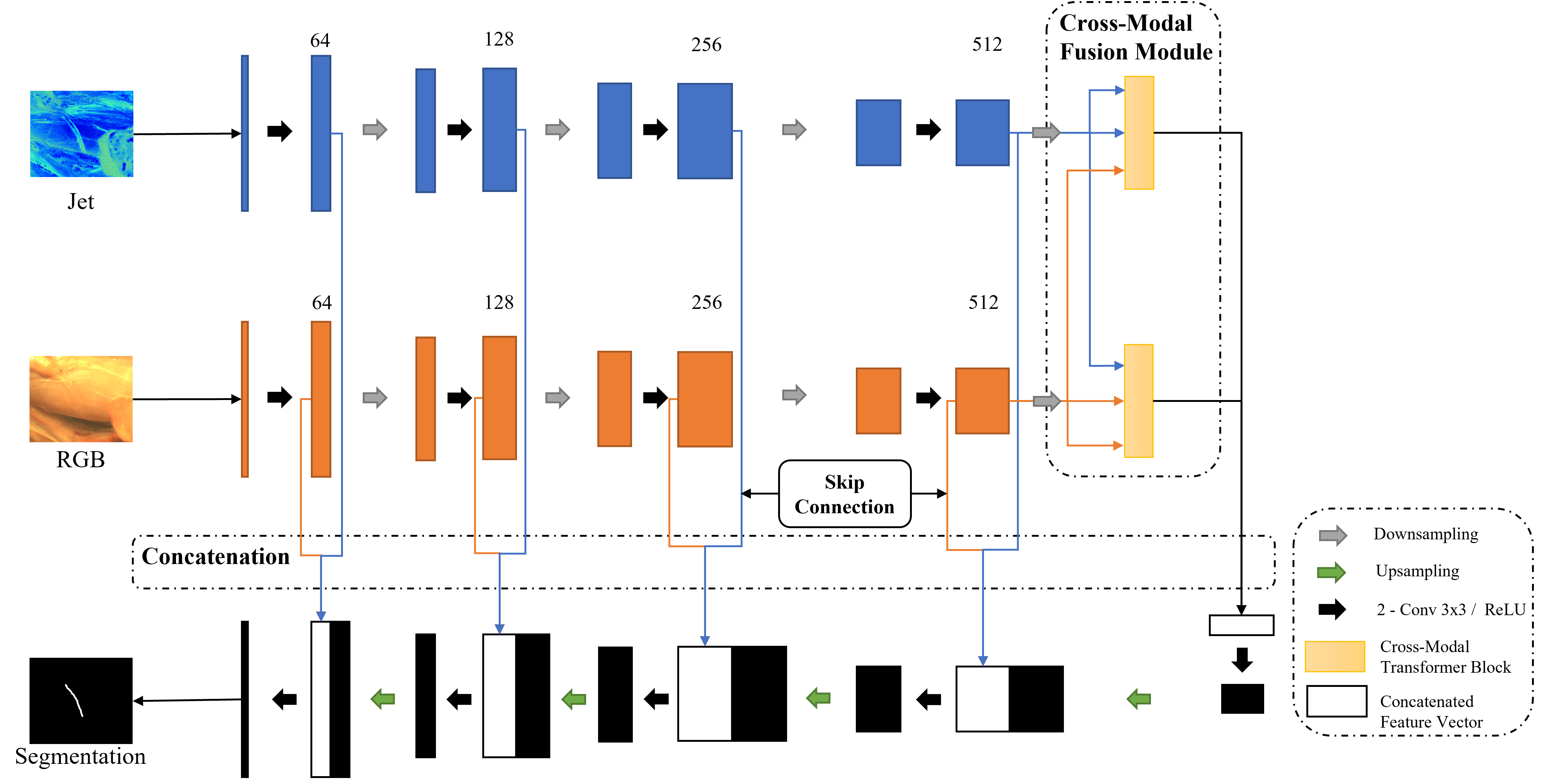}
	\caption{The block diagram of the Dual U-Net with the cross-modal Transformer block module.}
	\label{fig:model}
\end{figure*}

Our proposed network architecture is shown in Figure~\ref{fig:model}, in which two different modalities of the image are fused to learn the model. The parallel encoder structure allows the model to extract and combine the visual features of each independent modality. The base encoder is similar to the U-Net~\citep{ronneberger2015u}, which is aiming to generate and down-sample the feature map. Then we adapt the Transformer with the multi-head attention module proposed by \citet{vaswani2017attention} at the bottleneck. The multi-head attention module has been used for creating the self-attention feature map for image segmentation~\citep{petit2021u} but for a single modality. This study extends the idea and introduces the Transformer block for fusing different modalities of inputs via a cross-modal Transformer block. As can be shown in Figure~\ref{fig:crossmodal}, the expression of the attention output is given as follows:
\begin{equation}
    \textbf{Attention}(Q,K,V) = \textbf{softmax}(\frac{Q K^{^\top}}{\sqrt{d_k}})V,
\end{equation}
where $Q \in \mathbb{R}^{d \times d_B}$ is the queries from the modality A, $K \in \mathbb{R}^{d \times d_A}$ is the set of keys, and $V \in \mathbb{R}^{d \times d_A}$ is the set of values from the modality B.
The expression in the softmax function measures the similarity of $Q$ with respect to $K$. Then, the information from modality B is passed to modality A by calculating the products of attention function and $V$. Following the setting from the previous study~\citep{vaswani2017attention}, we add a residual connection and combine with the layer normalization and dropout operations. Furthermore, a feed-forward network with two hidden layers and another residual connection with the layer normalization and dropout are added for the Transformer block.

\subsection{Multi-task Learning}
A Loss function combined with an equal weight of the Binary Cross-Entropy (BCE) and Edge-Loss functions is used to optimize the model in this study.

The BCE is implemented with weight placed on positive examples to help with the imbalanced nature of the dataset. The BCE loss function with a positive weight is formulated as~\cite{ho2019real}:

\begin{equation}
    \ell_{(Y,P)} = - (1 / N) \sum_{n=1}^{N} w_p  y_{n}  log(p_{n}) + (1 - y_{n})  log(1 - p_{n}), 
\end{equation}

\noindent where $y_n$ $\in$ Y denotes the target labels and $p_n$ $\in$ P are the predicted probabilities for the $n^{th}$ pixel in the batch, with $N$ being the number of pixels in the batch. The Sigmoid activation function is used for the predicted probabilities. The weight placed on positive examples is denoted as $w_p$ in the loss function and calculated as follow:

\begin{equation}
    w_p = \frac{\textit{Total Background Pixels Of Set}}{\textit{Total Foreground Pixels Of Set}}.
\end{equation}

Positive weight is calculated on each individual training run and based on the ground truth masks of the current run's training set.

Following the definition of~\citet{xu2018cfun}, the Edge-Loss function is given by:

\begin{equation}
    \ell_{edge} = \frac{1}{C} \frac{1}{K} \frac{1}{N}\sum_{c=i}^C\sum_{k=1}^K\sum_{n=1}^N \left\|{E_k(y)^c_n - E_k(p)^c_n}\right\|_2 ,
\end{equation}
where $C$ is the classes of segmentation, $K$ is the number of Sobel kernels, $N$ is the pixel  number of the image, $y$ is the label map and $p$ is the prediction map. The final loss function to optimize the model is the sum of the BCE and Edge-Loss functions:
\begin{equation}
    \ell_{total} = \ell_{(Y,P)} + \ell_{edge}.
\end{equation}

\subsection{Training Implementation}

The Stochastic gradient descent (SGD) optimizer with a multi-step decayed learning rate is employed to train the network architectures. An initial learning rate of 0.03 decayed by 1/3 at 0.25 and 0.75 of the total number of epochs (250) is implemented.

Eight images of each modality are used for each mini-batch. Code implementation was performed in PyTorch, and experiments ran on four Nvidia Tesla V100 SXM2 16GB GPUs.

\begin{table*}[hbt!]
\centering
\caption{Validation result comparisons of different network structures for single modality.}
\resizebox{\textwidth}{!}{%
\begin{tabular}{|l|l|l|| c|c|c|c|c ||l|l|}
\hline
\multicolumn{3}{|l||}{} &
  \multicolumn{5}{l||}{Detection Metrics [Mean $\pm$ Standard Deviation \% ]} &
  \multicolumn{2}{l|}{Segmentation Metrics} \\ \hline
  \hline
\multicolumn{2}{|l|}{Image Representation} &
  CNN &
  \multicolumn{1}{l|}{Accuracy} &
  \multicolumn{1}{l|}{Sensitivity} &
  \multicolumn{1}{l|}{Specificity} &
  \multicolumn{1}{l|}{Precision} &
  \multicolumn{1}{l||}{Balanced Accuracy} &
  F2 &
  Dice \\ \hline
\multicolumn{2}{|l|}{{Jet}}  & U-Net      & $\mathbf{81.93}\pm$7.69 & $\mathbf{78.20}\pm$9.09 & 94.64$\pm$6.59 & 98.31$\pm$2.10 & $\mathbf{86.42}\pm$6.36 & 72.08$\pm$4.10 & 68.99$\pm$4.14 \\ \cline{3-10} 
\multicolumn{2}{|l|}{}       & Att. U-Net & 79.77$\pm$7.50 & 75.29$\pm$9.99 & 94.64$\pm$6.59 & $\mathbf{98.32}\pm$2.07 & 84.96$\pm$6.04 & 72.17$\pm$5.52 & 68.74$\pm$4.95 \\ \cline{3-10} 
\multicolumn{2}{|l|}{}       & Cross-Att. U-Net & 81.39$\pm$8.01 & 77.46$\pm$9.81 & 94.64$\pm$6.59 & 98.31$\pm$2.10 & 86.31$\pm$6.53 & $\mathbf{72.79}\pm$5.15 & $\mathbf{69.69}\pm$4.47 \\ \hline \hline

\multicolumn{2}{|l|}{{RGB}}  & U-Net      & 72.84$\pm$6.54 & 68.57$\pm$5.71 & $\mathbf{88.64}\pm$11.09 & 95.02$\pm$5.78 & $\mathbf{78.61}\pm$8.17 & 63.19$\pm$5.57 & 61.12$\pm$5.40  \\ \cline{3-10} 
\multicolumn{2}{|l|}{}       & Att. U-net & 72.35$\pm$2.03 & 67.91$\pm$3.58 & 87.92$\pm$8.04 & $\mathbf{95.31}\pm$2.69 & 77.92$\pm$3.69 & 63.37$\pm$3.02 & 60.94$\pm$3.41 \\ \cline{3-10} 
\multicolumn{2}{|l|}{}       & Cross-Att. U-net & $\mathbf{73.93}\pm$4.33 & $\mathbf{71.27}\pm$4.56 & 82.79$\pm$12.11 & 93.79$\pm$4.44 & 77.03$\pm$6.59 & $\mathbf{63.65}\pm$4.22 & $\mathbf{61.12\pm4.35}$ \\ \hline 
\end{tabular}%
}
\label{tab:singlevalresults}
\end{table*}

\begin{table*}[hbt!]
\centering
\caption{Validation result comparisons of different network structures for multi-modality.}
\resizebox{\textwidth}{!}{%
\begin{tabular}{|l|l|l|| c|c|c|c|c ||l|l|}
\hline
\multicolumn{3}{|l||}{} &
  \multicolumn{5}{l||}{Detection Metrics [Mean $\pm$ Standard Deviation \% ]} &
  \multicolumn{2}{l|}{Segmentation Metrics} \\ \hline
  \hline
\multicolumn{2}{|l|}{Image Representation} &
  CNN &
  \multicolumn{1}{l|}{Accuracy} &
  \multicolumn{1}{l|}{Sensitivity} &
  \multicolumn{1}{l|}{Specificity} &
  \multicolumn{1}{l|}{Precision} &
  \multicolumn{1}{l||}{Balanced Accuracy} &
  F2 &
  Dice \\ \hline
\multicolumn{2}{|l|}{{Jet + RGB}} & Dual U-Net & 85.63$\pm$5.95  & 82.92$\pm$7.27 & 94.64$\pm$6.59 & 98.45$\pm$1.90 & 88.78$\pm$5.26 & 74.40$\pm$4.46 & 71.09$\pm$4.67 \\ \cline{3-10} 
\multicolumn{2}{|l|}{}       & Co-Learn U-Net & 85.65$\pm$6.56 & 83.11$\pm$8.27 & 95.50$\pm$5.57 & 98.42$\pm$1.96 & 89.30$\pm$4.23 & 74.58$\pm$3.63 & 71.13$\pm$3.63 \\ \cline{3-10}
\multicolumn{2}{|l|}{}       & DXM-TransFuse U-net* & $\mathbf{88.34}\pm$7.51 & $\mathbf{85.95}\pm$9.21 & $\mathbf{97.50}\pm$5.00 & $\mathbf{99.29}\pm$1.43 & $\mathbf{91.72}\pm$4.73 & $\mathbf{76.12}\pm$3.40 & $\mathbf{72.10}\pm$3.99 \\ \hline
\end{tabular}%
}
\begin{tablenotes}
\item ($*$ is the proposed DXM-TransFuse U-net.)
\end{tablenotes}
\label{tab:multivalresults}
\end{table*}

\begin{figure*}[ht!]
    \centering
	\includegraphics[scale=0.60]{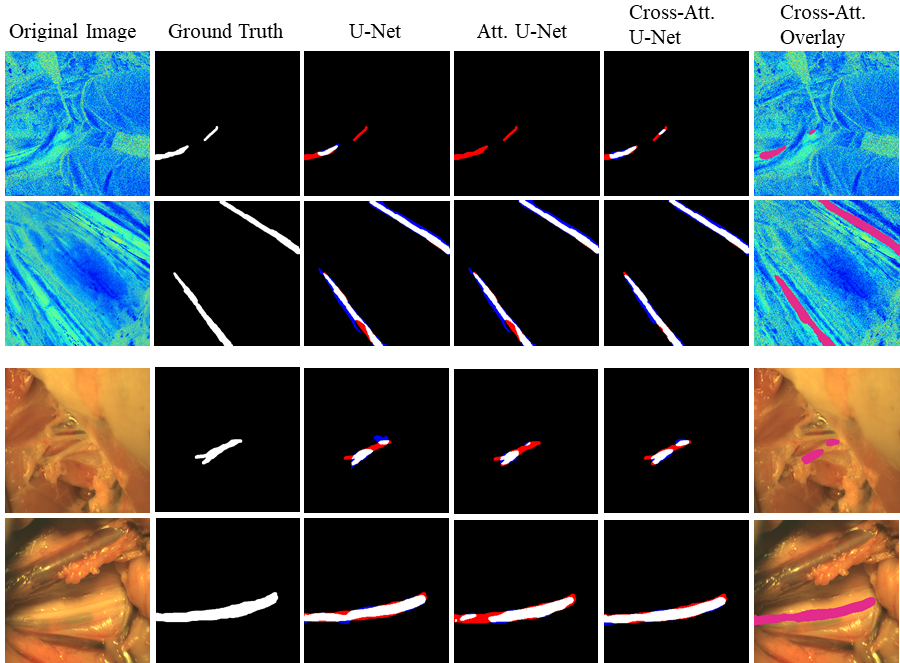}
	\caption{Predicted mask comparisons of single modality (Upper two rows: birefringence images, lower two rows: RGB images). }
	\label{fig:singlecombinedmaskcomparison}
\end{figure*}

\subsection{Evaluation}
To evaluate the model in this study, two types of metrics are used: one to evaluate the segmentation quality and the other to evaluate the detection of a nerve. A threshold of 0.5 is used for the predicted segmentation probabilities.

For evaluating segmentation quality, the two main metrics used are the dice coefficient and the F2 score, which are given as follows:
\begin{equation}
    \textit{Dice Coefficient} = \frac{\textit{TP} }{\textit{TP} + \frac{\textit{1}}{\textit{2}}( \textit{FP} + \textit{FN})},
\end{equation}

\begin{equation}
    \textit{F2 Score} = \frac{\textit{TP} }{\textit{TP} + \frac{\textit{1}}{\textit{5}}\textit{FP} + \frac{\textit{4}}{\textit{5}}\textit{FN}},
\end{equation}

Where TP is true positive, FP false positive, FN false negative and TN true negative.

Additionally, dice coefficient values are leveraged to help with the metrics used for the evaluation of nerve detection: accuracy, sensitivity/recall, specificity, precision, and balanced accuracy, which are given as follows:

\begin{equation}
    \textit{Accuracy} = \frac{\textit{Images with Dice} > \textit{Dice threshold}}{\textit{Total Images}},
\end{equation}

\begin{equation}
    \textit{Sensitivity / Recall} = \frac{\textit{TP} }{\textit{TP} + \textit{FN}},
\end{equation}

\begin{equation}
    \textit{Specificity} = \frac{\textit{TN} }{\textit{TN} + \textit{FP}},
\end{equation}

\begin{equation}
    \textit{Precision} = \frac{\textit{TP} }{\textit{TP} + \textit{FP}},
\end{equation}

\begin{equation}
    \textit{Balanced Accuracy} = \frac{\textit{Sensitivity} + \textit{Specificity}}{\textit{2}}.
\end{equation}

To help with the classification metric, labels are set as part of the dataset to determine if an image contains nerve tissues or not. If the dice value of the predicted mask is greater than the dice threshold, this is considered a true positive if it is labeled to contain nerve tissues; otherwise, it is considered a true negative. If the dice is below the threshold, it is considered a false negative if it is labeled as containing nerve tissues; otherwise, it is considered false positive. 


Results are based on the 5-fold cross-validation, where one fifth is the hold-out set used for validation with the remaining used for training.

\section{RESULTS} \label{results}

\begin{figure*}[ht]
    \centering
	\includegraphics[scale=0.60]{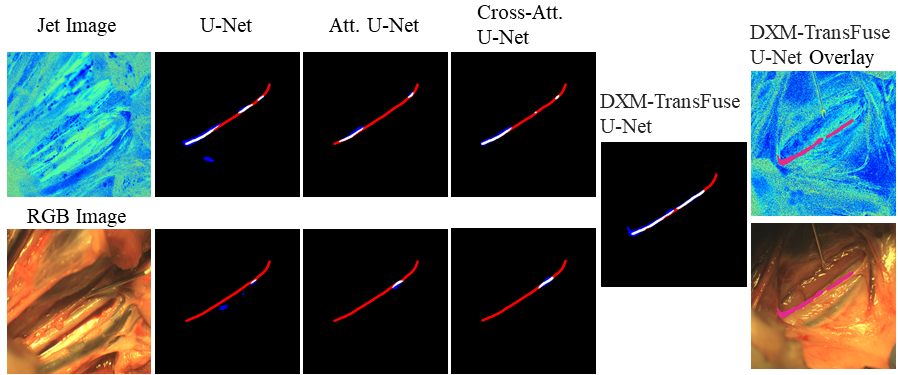}
	\caption{Predicted mask comparisons between single modality and the proposed multi-modality networks.}
	\label{fig:singledualcomp}
\end{figure*}

\begin{figure*}[ht]
    \centering
	\includegraphics[scale=0.60]{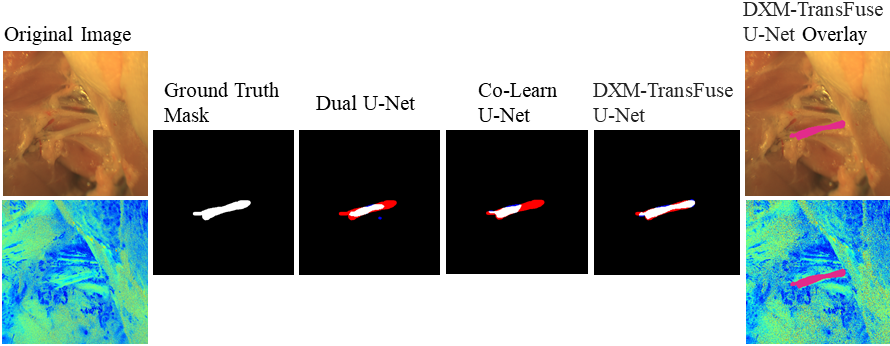}
	\caption{Predicted mask comparisons among different multi-modality networks.}
	\label{fig:dualcomp}
\end{figure*}

Table~\ref{tab:singlevalresults} provides the comparison of the validation results between the network structures for each image representation. Concerning the birefringence images, the cross-attention U-Net had the highest dice and F2 scores while U-Net had better nerve detection metrics, including the accuracy, sensitivity, and balanced accuracy. In regards to the RGB images, the cross-attention U-Net had higher accuracy, sensitivity, and F2 than the other models, the U-Net outperformed in specificity and balanced accuracy, and Attention U-Net showed the best in precision. When comparing between the image modalities, the models using birefringence images outperformed the ones with RGB images across all metrics. It is worth noting that due to resource limitations, only a single attention head was used in the cross-attention U-Net model.

Figures~\ref{fig:singlecombinedmaskcomparison}, \ref{fig:singledualcomp} and \ref{fig:dualcomp} demonstrate the mask differences between ground truth and the predicted mask of its respective network. On the mask comparison images, sections highlighted in \textit{blue} are related to an over-segmentation prediction, \textit{red} where under segmentation occurred, and \textit{white} where correctly segmented.

Figure~\ref{fig:singlecombinedmaskcomparison} illustrates the differences between the single modality U-Net, the Attention U-Net, and the Cross Attention U-Net approaches. The upper row demonstrates a particularly difficult case with a birefringence image where it can be seen that the Attention U-Net was not able to detect the nerve, the U-Net had a lot of under segmentation, but the Cross Attention U-Net was able to detect more segments of the nerve. The second row shows that in cases where all three models performed well with a birefringence image, the Cross Attention still outperformed the others with minimal under-segmentation. The third row on Figure~\ref{fig:singlecombinedmaskcomparison} demonstrates a challenging case with an RGB image where the Cross Attention U-Net had correctly detected more portion of the nerve with less over-segmentation than U-Net, while all other architectures had some under-segmentation. The last row further demonstrates that the Cross Attention minimized the under segmentation error in cases where other models performed well with an RGB image. In general, introducing a Cross Attention module is shown to be advantageous for extracting salient features and suppressing irrelevant information.

Table~\ref{tab:multivalresults} provides the validation comparison between the dual modality networks. All three Multi-modality networks had significantly outperformed single-modality networks across all metrics. Furthermore, within the multi-modality networks, the proposed DXM-TransFuse U-net had the best detection and segmentation metrics.

Figure~\ref{fig:singledualcomp} demonstrates that the proposed DXM-TransFuse U-net is able to leverage each individual modality's information to improve the overall nerve detection. As can be seen, all single modality networks, regardless of image representation was unable to detect the nerve correctly. However, the DXM-TransFuse U-net was able to improve the detection in such a case drastically.

\begin{table}[htb]
\caption{Network Specific Details}
\resizebox{\columnwidth}{!}{%
\begin{tabular}{|l||l|l|}
\hline
  CNN &
  Inference Time (ms) &
  Number of Parameters \\ \hline \hline
U-Net & 17.76 $\pm$ 0.02 & 34,527,041 \\ \cline{1-3}
Att. U-Net & 19.28 $\pm$ 0.04 & 34,878,573  \\ \cline{1-3} 
Cross-Att. U-Net & 49.18$\pm$0.14 & 37,324,801  \\ \hline \hline
Dual U-Net      & 27.07$\pm$0.04 & 47,068,289  \\ \cline{1-3} 
Co-Learn U-Net & 29.68$\pm$0.01 & 56,506,497  \\ \cline{1-3} 
DXM-TransFuse U-Net & 29.30$\pm$0.01 & 53,373,057 \\ \hline 
\end{tabular}%
}
\label{tab:networkdetails}
\end{table}

Figure~\ref{fig:dualcomp} illustrates the differences between the predicted masks of each multi-modal network. It can be seen that the proposed DXM-TransFuse U-net had shown greater performances in the segmentation task. Both the Dual U-Net and Co-Learn U-Net networks showed a lot of under-segmentation, while the DXM-TransFuse U-net only slightly missed small portions of the nerve.

\begin{table*}[hbt!]
\centering
\caption{Test result comparisons of different network structures for single and multi modality}
\resizebox{\textwidth}{!}{%
\begin{tabular}{|l|l|l|| c|c|c|c|c ||l|l|}
\hline
\multicolumn{3}{|l||}{} &
  \multicolumn{5}{l||}{Detection Metrics [Mean $\pm$ Standard Deviation \% ]} &
  \multicolumn{2}{l|}{Segmentation Metrics} \\ \hline
  \hline
\multicolumn{2}{|l|}{Image Representation} &
  CNN &
  \multicolumn{1}{l|}{Accuracy} &
  \multicolumn{1}{l|}{Sensitivity} &
  \multicolumn{1}{l|}{Specificity} &
  \multicolumn{1}{l|}{Precision} &
  \multicolumn{1}{l||}{Balanced Accuracy} &
  F2 &
  Dice \\ \hline
\multicolumn{2}{|l|}{{Jet}}  & U-Net      & 84.10$\pm$2.51 & 84.24$\pm$2.97 & 83.33$\pm$0.00 & 96.25$\pm$0.12 & 83.79$\pm$1.48 & 71.88$\pm$1.98 & 67.10$\pm$1.31 \\ \cline{3-10} 
\multicolumn{2}{|l|}{}       & Att. U-Net & 83.08$\pm$2.05 & 83.03$\pm$2.42 & 83.33$\pm$0.00 & 96.48$\pm$0.10 & 83.18$\pm$1.21 & 71.20$\pm$0.63 & 66.80$\pm$0.66 \\ \cline{3-10} 
\multicolumn{2}{|l|}{}       & Cross-Att. U-Net & $\mathbf{85.13}\pm$1.03 & $\mathbf{85.45}\pm$1.21 & 83.33$\pm$0.00 & 96.57$\pm$0.05 & 84.39$\pm$0.61 & 71.99$\pm$1.16 & 67.31$\pm$0.80 \\ \hline \hline
\multicolumn{2}{|l|}{{Jet+ RGB}}  & Dual U-Net      & 83.08$\pm$3.08 & 84.24$\pm$2.97 & 76.67$\pm$13.33 & 95.28$\pm$2.48 & 80.45$\pm$6.68 & 71.65$\pm$1.22 & 66.80$\pm$2.62  \\ \cline{3-10} 
\multicolumn{2}{|l|}{}       & Co-Learn U-Net & 81.54$\pm$1.03 & 81.21$\pm$1.21 & 83.33$\pm$0.00 & 96.40$\pm$0.05 & 82.27$\pm$0.61 & 72.63$\pm$1.14 & 67.47$\pm$0.79 \\ \cline{3-10} 
\multicolumn{2}{|l|}{}       & DXM-TransFuse U-Net & 84.62$\pm$2.29 & 84.24$\pm$2.27 & $\mathbf{86.67}\pm$6.67 & $\mathbf{97.22}\pm$1.40 & $\mathbf{85.45}\pm$3.66 & $\mathbf{73.97}\pm$1.65 & $\mathbf{68.35}\pm$0.93 \\ \hline 
\end{tabular}%
}
\label{tab:testresults}
\end{table*}

As shown in Tables~\ref{tab:singlevalresults} and~\ref{tab:multivalresults}, the F2 score, which places more emphasis on under segmentation, would be a better predictor of the model performance with respect to nerve detection as it is more concerning if we incorrectly detected part of the nerve than over-estimated the area of the nerve. Considering all these results, the proposed DXM-TransFuse U-net has the highest performance for multi-modality, with the Cross-Attention U-Net also showing effectiveness for single-modality.

\section{CONCLUSIONS}

In this study, we proposed the DXM-TransFuse U-net that could fuse information from multi-modal medical images. The findings of the study showed that introducing the Cross Attention on a single modality network improved the detection and segmentation of nerve structures on the birefringence images. Furthermore, by employing parallel encoders to extract the feature maps of each modality independently and applying the Transformer module to fuse the information of each, the network with Transformer was able to improve further its segmentation of nerve structures resulting in better overall detection. However, given the relatively small sample size in this study due to the difficulty in acquiring a large dataset of clinical nerve images, challenges were present in the development and validation of our models. To ensure the robustness of the models, we applied the 5-fold validation during the experiments. Although data augmentation was applied in this study, the model's generalization can be enhanced further by increasing the dataset in future studies. The current multi-modality framework hit the limits of our computational resources, but we plan to combine more modalities of inputs for training the model in our future work. Nonetheless, our research will be constructive in combining different image modalities for nerve image segmentation.

Additional modifications to the network can be implemented by extending the usage of the Transformer module at each skip connection layer to strengthen the cross-modal interactions between the modalities. Due to the differences in the variance in brightness and contrast of each imaging modality, further investigations to determine optimal color augmentation criteria for each modality are recommended for the future studies.



\section*{ACKNOWLEDGMENT}
This work was partially supported by Sheikh Zayed Institute for Pediatric Surgical Innovation through the Smart Tools program (PI: Cha, SPF44215-30005967) and Children’s National Board Of Visitors Grant Award (PI: Cha, 44235-60006539) and by the George Washington University SEAS Startup fund (PI: Park). Authors would like to thank Neurovision Medical Products for a loaner device and research support. The authors would also like to appreciate Dr. Anthony D. Sandler for the preclinical study support and his advice on clinical insights.


\bibliographystyle{model5-names}
\biboptions{authoryear}
\bibliography{root}





\end{document}